# Second order add/drop filter with a single ring resonator

Matteo Cherchi*, Fei Sun, Markku Kapulainen,
Tapani Vehmas, Mikko Harjanne, and Timo Aalto
VTT Technical Research Centre of Finland, Tietotie 3, 02150 Espoo, Finland

## ABSTRACT

We show theoretically and experimentally how a flat-top second-order response can be achieved with a self-coupled single add-drop ring resonator based on two couplers with different splitting ratios. The resulting device is a 1x1 filter, reflecting light back in the input waveguide at resonating wavelengths in the passbands, and transmitting light in the output waveguide at all other non-resonating wavelengths. Different implementations of the filter have been designed and fabricated on a micron-scale silicon photonics platform. They are based on compact Euler bends - either U-bends or L-bends - and Multi-Mode Interferometers as splitters for the ring resonators. Different finesse values have been achieved by using either 50:50 MMIs in conjunction with 85:15 MMIs or 85:15 MMIs in conjunction with 95:05 double MMIs. Unlike ordinary lowest order directional couplers, the MMIs couple most of the power in the cross-port which make them particularly suitable for the topology of the self-coupled ring, which would otherwise require a waveguide crossing. Experimental results are presented, showing good agreement with simulations. The proposed devices can find applications as wavelength-selective reflectors for relatively broad-band lasers or used as 2x2 add-drop filters when two exact replicas of the device are placed on the arms of a Mach-Zehnder interferometer.

## 1. INTRODUCTION

Optical filters with flat-top response are highly desirable and often mandatory, especially in telecommunication applications, where in-band ripples can cause unwanted signal distortions (e.g. in high-speed links) and/or power variations induced by wavelength drifts of the light sources (e.g. in CWDM systems with no thermal control). One of the most popular solutions for optical filtering is the ring-resonator[1,2], which is the integrated-optics version of the Fabry-Perot interferometer. Given that a single ring resonator inherently shows the typical Lorentzian response of Fabry-Perot resonators, the typical approach to achieve flat-top response is to combine multiple ring resonators, either in series[3] or in parallel[4]. Configurations with two coupled resonators (i.e. second-order filters) have been also exploited to highlight Fano resonances[5] (often referred as all-optical analogue to electromagnetically induced transparency) in optical circuits[6]. More recently a few papers have shown that Fano resonances can be obtained also in a single ring resonator, by suitably coupling light from more than one port[7,8]. Remarkably, all configurations reported to date require a waveguide crossing when implemented in a photonic integrated circuit (PIC).

In the present paper we focused on the simple configuration proposed by Wang et al.[8], with two main goals: first to demonstrate that it is possible to design a topologically equivalent structure without any crossing, and second that a suitable design of the coupling ratios allow to achieve flat-top response instead of a Fano resonance that is of little use for real applications.

The paper is organized as follows: in section 2 the novel configuration with no crossing is introduced, in section 3 the generic implementation of ring resonators in VTT micron-scale silicon photonic platform is explained, in section 4 the design of flat top filters is reported, in section 5 the layout and fabrication of the filters is presented, and in section 6 the experimental results are discussed. Conclusions and perspectives are eventually provided.

## 2. A SECOND-ORDER RING RESONATOR WITHOUT CROSSINGS

A very simple second order response from a single-ring resonator has been achieved by Wang et al.[8] by connecting the two couplers of the ring through a waveguide, so to create two counter-propagating light paths in the same ring. The resonating light is back-reflected in the input port, effectively acting also as drop port, whereas non-resonating light is transmitted in the output port, acting as through port. Assuming the two couplers of the ring resonator are identical, at exact resonance the interplay of the two coupled counter-propagating modes induces destructive interference in the drop port, resulting in a narrow peak in the middle of the resonant dip of the through port. As shown in Fig. 1, the coupling configuration requires a crossing of the output port with the connecting waveguide (blue). This was not a big deal in the original paper, where the concept has been demonstrated based on optical fiber loops, but it can be an issue in an integrated version in a PIC. In fact, even if waveguide crossings are possible in most waveguide platforms, crossing losses are usually non-negligible, special fabrication steps are typically required, and the footprint of the crossing can also be large. Actually crossing losses have little impact in this case, as the crossing does not affect the ring resonator and its quality factor, but the requirement of additional etch-steps, together with the relatively larger footprint, have bad impact on the fabrication cost.

*silicon.photonics@vtt.fi; phone +358 40 684 9040; fax +358 20 722 7012; http://www.vtt.fi/siliconphotonics

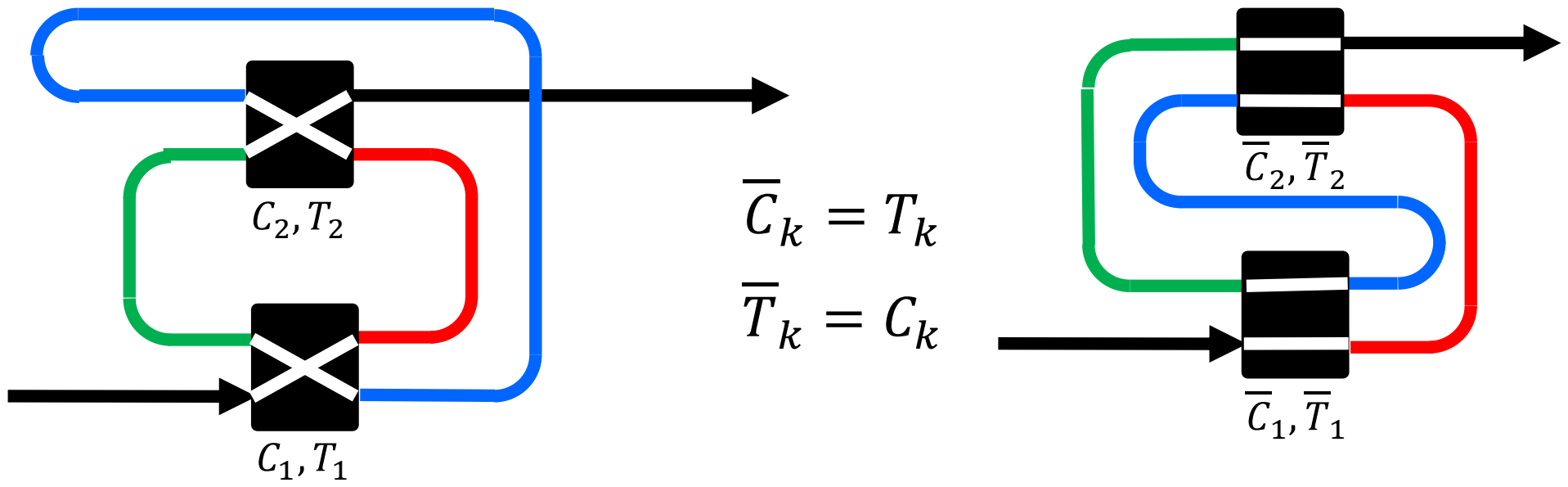

Fig. 1. Schematic representation of the resonator proposed by Wang et al.[8] (left) and proposed version with no crossings (right), based on complementary coupling ratios.

In order to avoid crossings, we propose to replace the original couplers (having coupling $C_1$ and $C_2$) of the ring resonator with complementary couplers having inverse splitting ratio (i.e. $\bar{C}_1 = 1 - C_1$ and $\bar{C}_2 = 1 - C_2$), and reconnect the waveguides in order to ensure the same coupling to all paths, as shown in Fig. 1. This way, the connecting waveguide will be embedded inside the ring, and no crossing is required anymore. In a sense, we can claim that the crossing is effectively moved inside the couplers. To better understand the topological equivalence of the two configurations, in Fig. 2 we have highlighted rings (pink) and connectors (yellow) both in the original configuration and in our upgraded version.

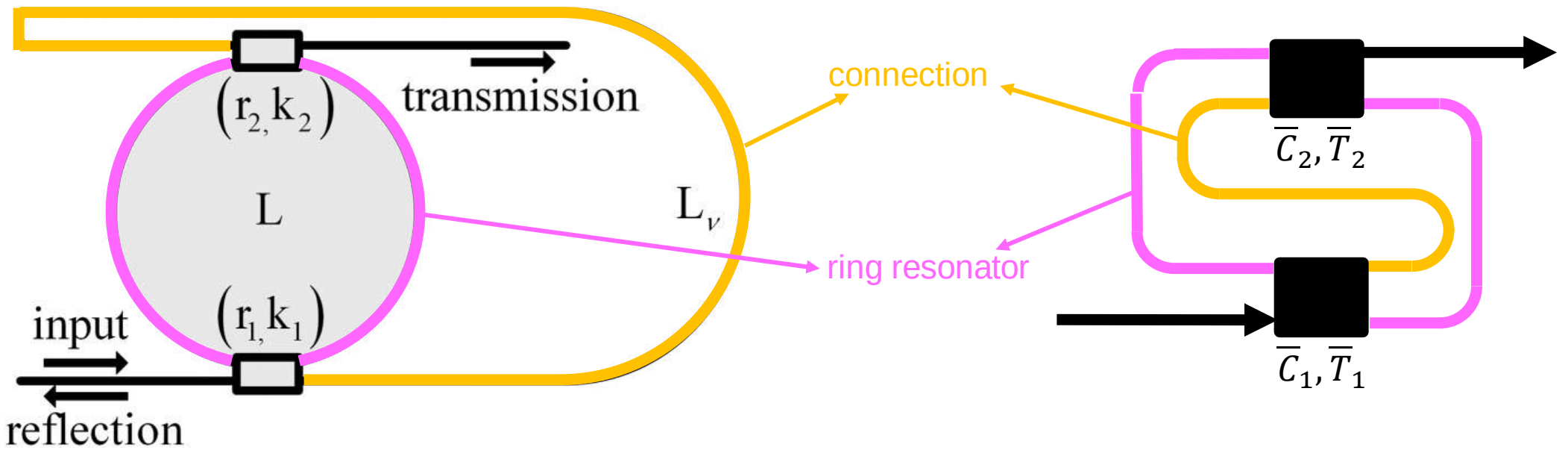

Fig. 2. Direct comparison of the configuration by Wang et al.[8] (left) and the novel configuration (right), in order to highlight the ring resonator (pink) and the connecting waveguide (yellow).

It may be argued that, in most practical cases, the original coupling of the couplers is small, whereas our version requires high coupling, which is not easy to achieve in most platforms with directional couplers, unless long couplers are used, with potential bad impact on the size of the ring resonator and/or on its losses. Another issue with our version is that the ring itself is broken into two pieces divided by the couplers, which means that transition losses in the couplers, usually negligible, can instead have a non-negligible impact on the quality factor of the resonator. Even if these argument make a lot of sense in most waveguide platform, they don't actually apply to VTT platform, where ring resonators are made based on multi-mode interference (MMI) splitters as couplers, which, unlike normal directional couplers, have most of the power coupled in the cross port instead of the bar port. This will be explained in more details in the next section.

# 3. RING RESONATORS IN VTT PLATFORM

VTT platform is an interesting alternative silicon photonics platform, based on 3 µm thick silicon on insulator (SOI) wafers. Single mode condition is ensured by suitably designed rib waveguides[9], obtained by etching 40% of the silicon layer thickness. Strip waveguides are obtained by etching the remaining silicon through a self-aligned process[10], also ensuring very low loss rib-to-strip converters[11]. Total internal reflection (TIR) in silicon can also be exploited to design and fabricate low loss mirrors for both rib and strip waveguides[12]. All these elements are schematically depicted in Fig. 3.a. A thermal oxidation step is applied to all types of waveguides, leading to propagation losses as low as 0.10 dB/cm in the rib waveguides and 0.13 dB/cm in the strip waveguides. Beside TIR mirrors, micron-scale bends are also possible in multimode strip waveguides, which effectively operate as single-mode, thanks to a recent breakthrough[13,14] (see Fig. 3.b), which led also to a patent[15].

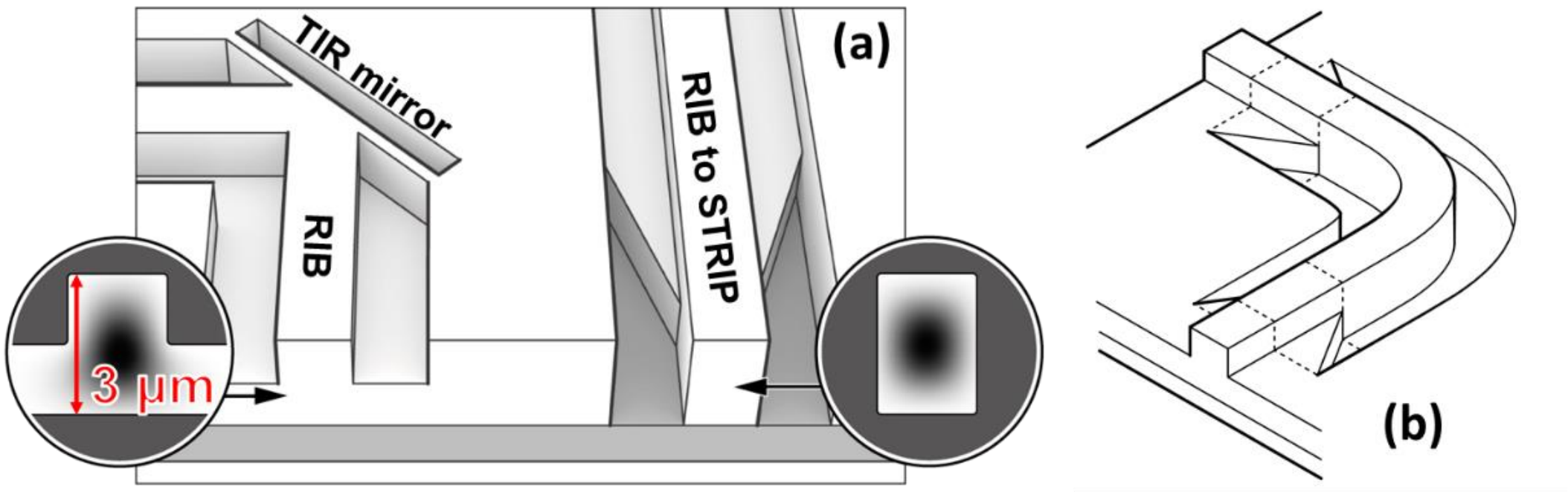

Fig. 3. Exemplified schematics of the VTT silicon photonics platform. a) Rib and strip waveguides with their mode distribution, rib to strip converter, and TIR mirror; b) the Euler bend.

In principle there are two approaches to implement ring resonators in VTT platform. Implementations based on rib-waveguides are limited by the large bending radius (about 5 mm) required and by the length of the couplers. Furthermore rib couplers are very sensitive to etch-depth variations of the rib waveguides, which are typically around ± 5% at wafer scale, making the Q to sensibly change all over the wafer and from wafer to wafer. The most promising approach is instead based on strip waveguides and Euler bends, making sure that higher order modes are not excited.

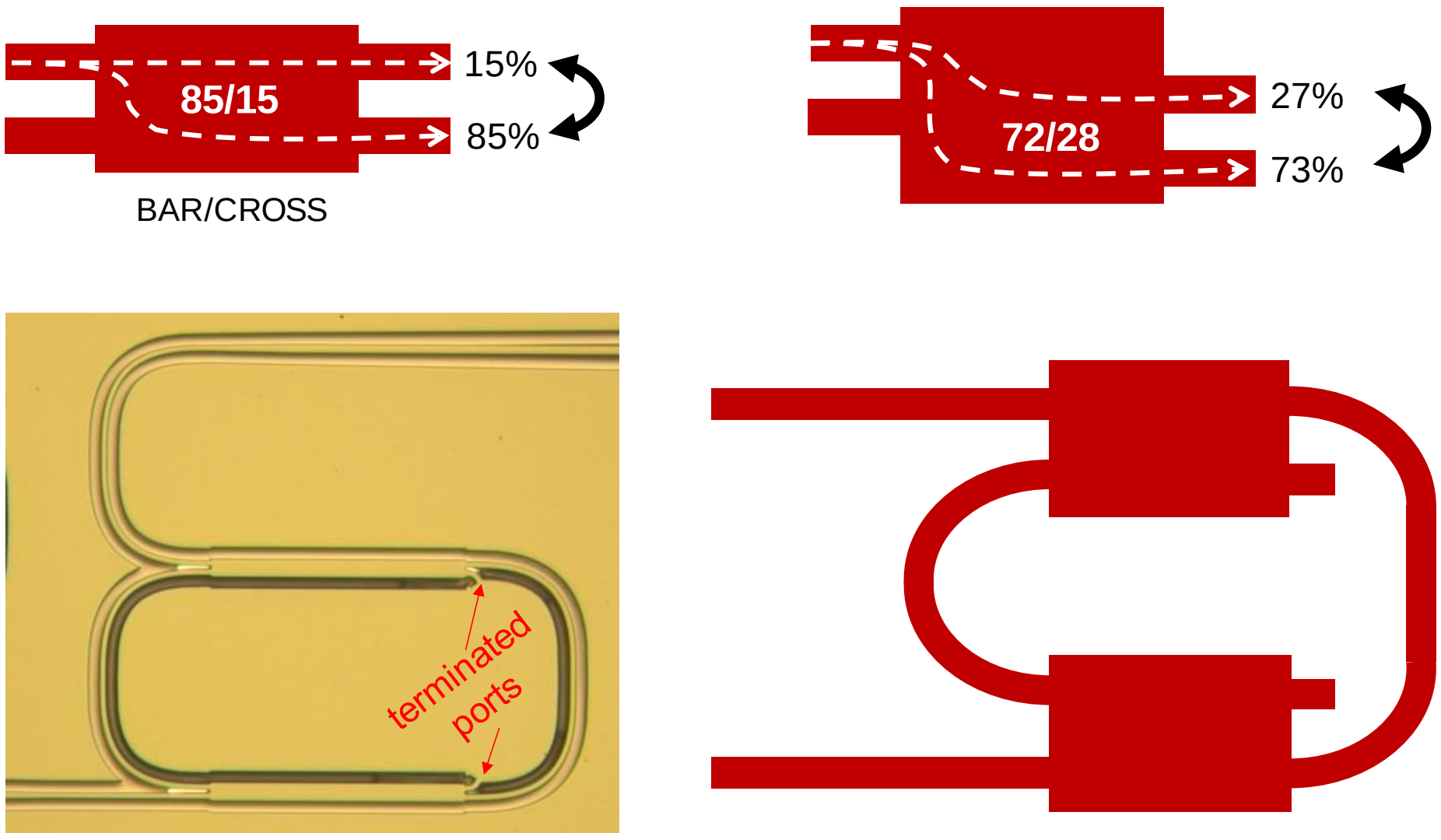

Fig. 4. Ring resonators based on standard MMI splitters: only two ports are accessible.

But a major problem with this approach is that light is very well confined inside the strip waveguide, which means that, taking also into account the lithographic limitations for minimum achievable gap sizes, directional couplers are not an option. The only viable way to couple light between this type of waveguides is using MMI splitters. MMIs are well known for their relatively broadband operation and tolerance to fabrication errors, the only limitation being that covering all possible splitting ratios is not trivial. In this paper we tried to use as much as possible the discrete set of splitting ratios available in standard MMI splitters, and resort to double MMIs[16] only when strictly necessary, so to minimize the number of elements in the filter layout. In principle also tapered MMIs[17] could have been used instead, to shorten the length of the resonators.

Ring resonators have been fabricated on VTT platform based on standard MMIs that, unlike usual directional couplers, have most of the power coupled in the bar port, as depicted in Fig. 4. This results in an odd configuration, where two of the ring ports, namely the through and add ports, are not available, as they are embedded inside the ring. The only way to access them would be through waveguide crossings that would affect the ring quality factor. A typical response from the drop port of a MMI based on 85:15 MMIs is reported in Fig. 5, with finesse 16 and quality factor 12,000. In the left-hand side of all peaks it can be seen a significant distortion of the Lorentzian peak, most likely due to excitation of higher-order modes.

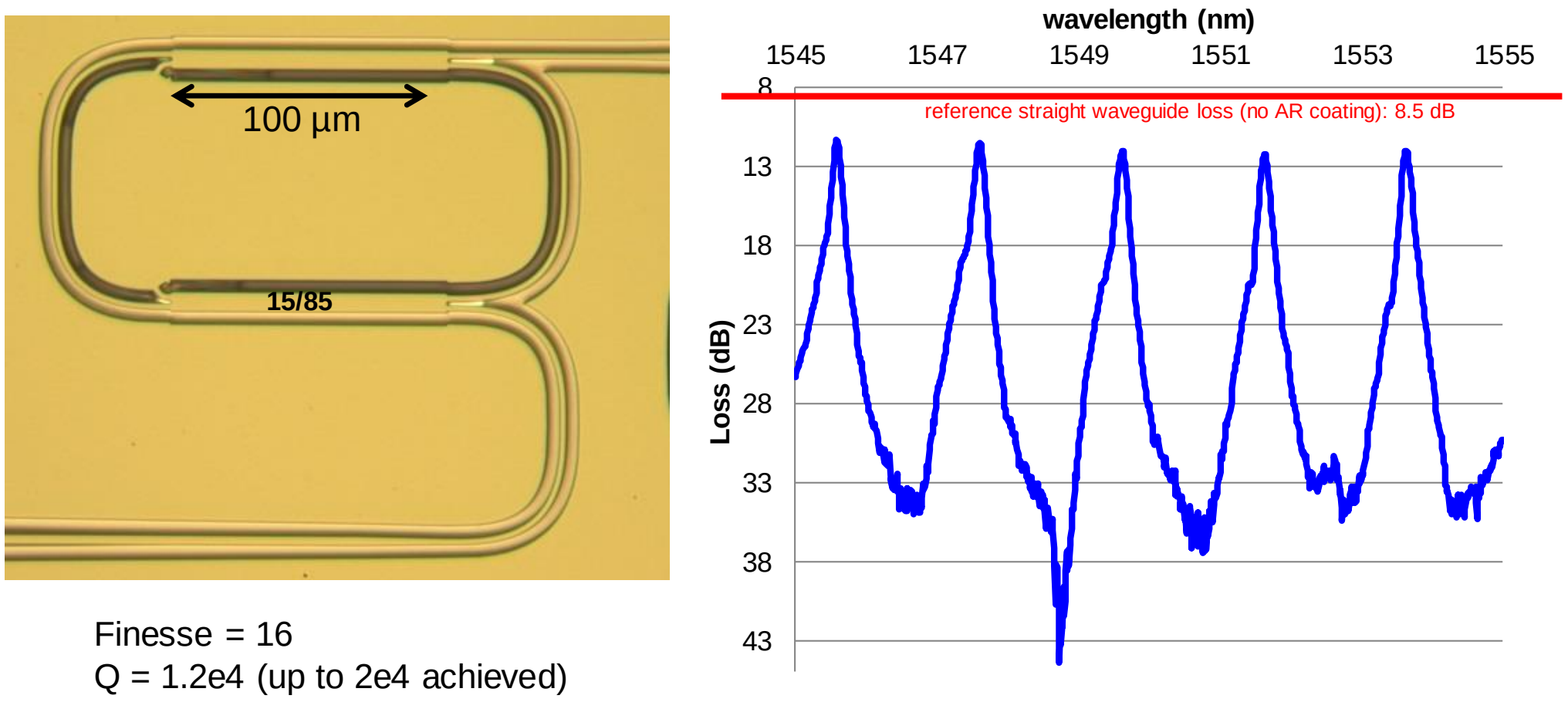

Finesse = 16
Q = 1.2e4 (up to 2e4 achieved)
FSR = 2.1 nm

Fig. 5. Example spectrum of a fabricated ring based on 85:15 MMIs and Euler bends.

We notice that, whereas having most of the power in the bar port is a major limitation in standard ring resonators, the analysis in section 2 clearly showed that the very same property is instead an advantage for the particular configuration under study in this paper, as it avoids using any crossing. In other words, we can claim that our platform is perfectly suitable to implement the type of second order filter in Fig. 2, thanks to the splitting properties of the MMIs.

## 4. DESIGN OF FLAT-TOP FILTERS

The amplitude transmission $\tau$ and reflection $\rho$ of the filter in Fig. 2 can be expressed as[8]

$$\tau = \left(ae^{i\delta}\right)^{\nu} \frac{\left(t_1 - t_2 ae^{i\delta}\right)\left(t_2 - t_1 ae^{i\delta}\right) + c_1^2 c_2^2 ae^{i\delta}}{\left(1 - t_1 t_2 ae^{i\delta}\right)^2} \qquad (1)$$

and

$$\rho = -2\left(ae^{i\delta}\right)^{v} \frac{\left(t_1 - t_2 ae^{i\delta}\right) c_1 c_2 \sqrt{ae^{i\delta}}}{(1 - t_1 t_2 ae^{i\delta})^2} \ , \tag{2}$$

where $t_1 \equiv \sqrt{T_1}$, $t_2 \equiv \sqrt{T_2}$, $c_1 \equiv \sqrt{C_1}$, and $c_2 \equiv \sqrt{C_2}$ are the bar and cross coupling coefficients of the original design, i.e. the cross and bar coefficients of the novel proposed layout. The parameter $\delta$ accounts for the phase accumulated in a round trip in the ring, while $a$ is the amplitude attenuation per round trip. The parameter $v$ is the ratio of the connector length relative to the ring length.

Based on these formulas, we have explored a few different configurations leading to flat-top response, trying to exploit as much as possible standard MMI splitting ratios. The ideal spectral response of the two chosen configurations are shown in Fig. 6, corresponding to completely lossless devices.

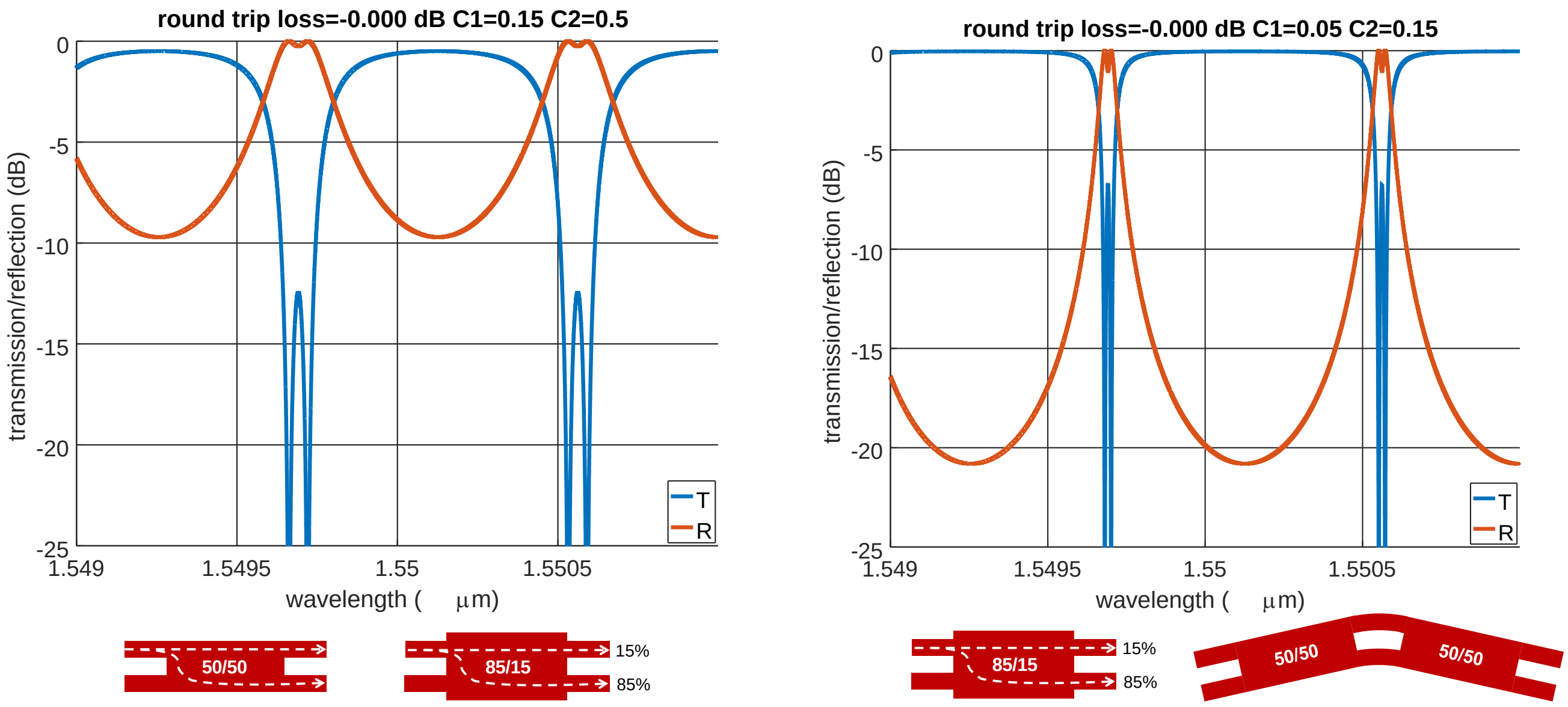

Fig. 6. Simulated spectral response of the chosen filters.

By choosing significantly different splitting ratio for the two MMIs of the ring the Fano resonance can be significantly (but not completely) suppressed, resulting in a flat-top response. The first combination has low finesse, as splitting ratios are 50:50 and 85:15. This is clear also from the low contrast (< 10 dB) of the resonances of the drop port (reflection R in the plots). The second choice has significantly higher finesse, being based on 85:15 and 95:05 splitting ratio. The last one is obtained by cascading two 50:50 MMI connected by two waveguides with a small tilt (see schematic picture in Fig. 6). The impact of losses, especially the ones coming from the MMI splitters, will be clearly higher on the high finesse version. Typical measured losses from a MMI splitter are between 0.1 and 0.2 dB, which means that the maximum round trip loss could be estimated around 0.5 dB in the worst case. A more detailed analysis of the impact of losses will be presented in section 6.

## 5. LAYOUT AND FABRICATION

Four different designs have been included in the mask layout, with two different finesse values and either L-bends or U-bends, as illustrated in Fig. 7. The input/output waveguides were designed to match a fiber array polished with 8° angle, in order to make characterization easier and suppress residual back-reflections from the silicon facet to the fiber. The rib waveguides are tapered up to 13 µm width at the interface, so to optimize coupling with the fiber modes.

We have fabricated the designed structures on 3 µm thick SOI wafers based on the well assessed VTT silicon waveguide technology. The devices were fabricated using smart-cut silicon-on-insulator wafers from SOITEC.

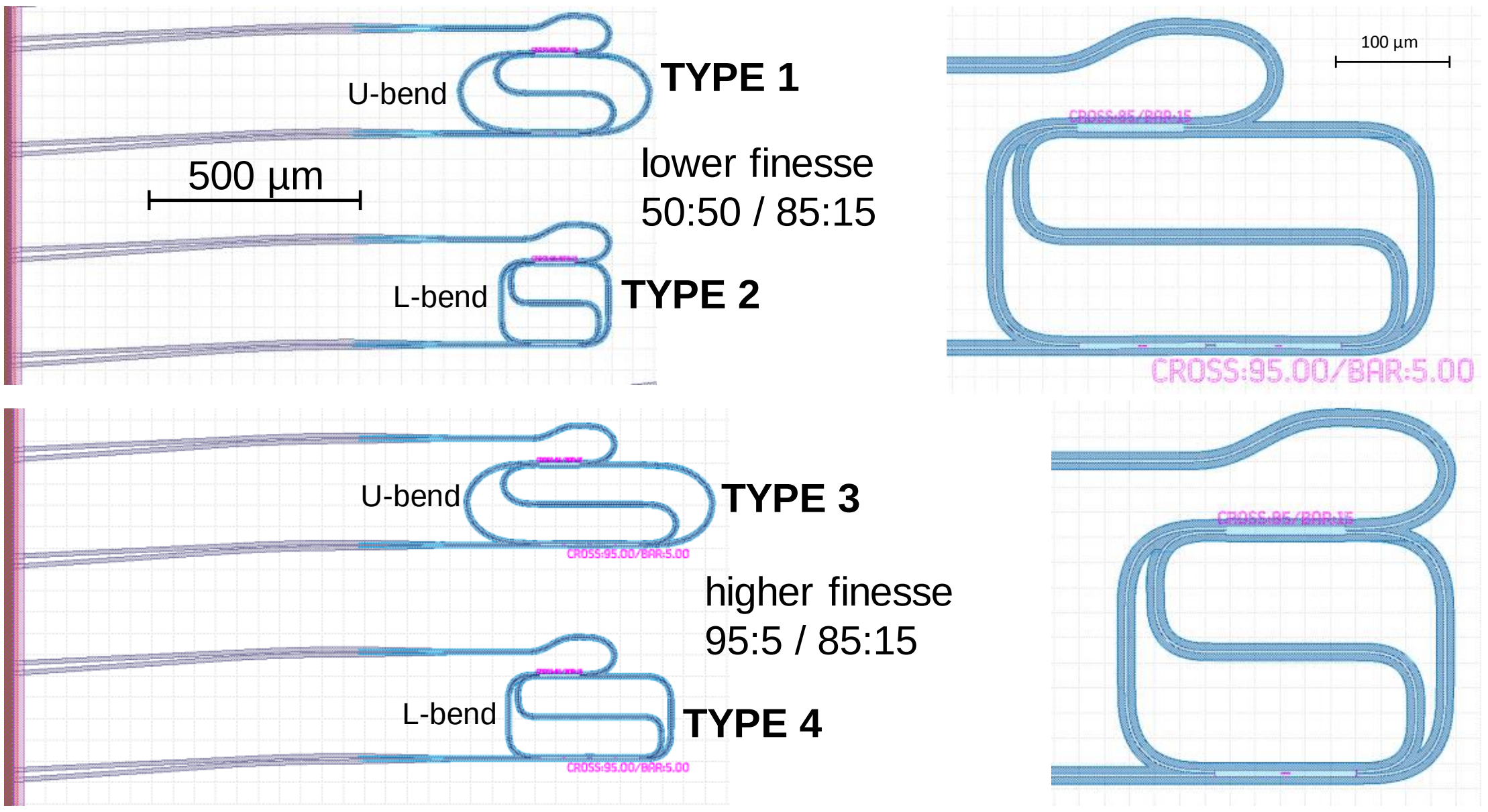

Fig. 7. The four types of filters put on mask (left) and detail of the two implementations based on L-bends, namely Type 4 and Type 2.

Initial SOI layer thickness was increased with epitaxial silicon growth. FilmTek 4000 spectrophotometer was used to measure the thicknesses of the SOI layer and the underlying buried oxide. A 500 nm thick Tetraethyl orthosilicate (TEOS) layer was deposited on the wafer in LPCVD diffusion furnace to work as a hard mask in silicon etching. Waveguide fabrication was done using our standard double-masking multi-step process[8]. In the process, two mask layers are passively aligned with respect to each other, and two separate silicon etch-steps form the rib and strip waveguide structures and waveguide facets. Lithography steps were done using FPA-2500i4 i-line wafer stepper from Canon Inc. and pattern transfer to oxide hard mask was done using LAM 4520 reactive ion etcher with $CF_4$ and $CHF_3$ chemistry. Waveguides were etched into silicon using Omega i2l ICP etcher from SPTS Technologies. The etching was done with a modified Bosch process[21] using $SF_6$ and $C_4F_8$ as etch and passivation gases, respectively, and $O_2$ as an etch gas to break passivation polymer formed by $C_4F_8$. After silicon etching, TEOS hard mask was removed with buffered oxide etch (BOE). Hard mask removal was followed by wet thermal oxidation consuming 225 nm of silicon, and thermal oxide removal with BOE. This was done to smoothen the etched surfaces, and to thin the SOI-layer to its final thickness of approximately 3 µm.

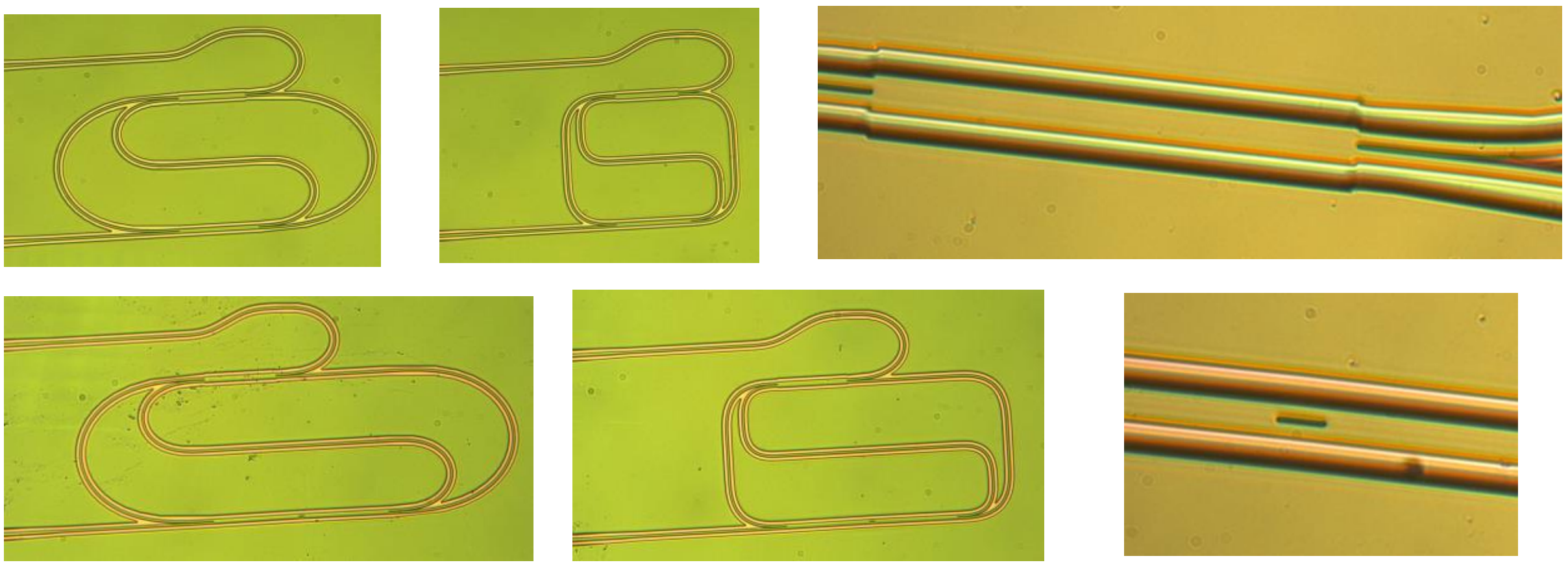

Fig. 8. Micrographs of some of the fabricated filters and details of a 85:15 MMI splitter and of the slightly bent junction between two 50:50 MMIs in a 95:05 double MMI.

A 0.17 µm thick silicon nitride layer was then deposited on the wafer with LPCVD as an antireflection coating and to prevent the etching of buried oxide in later oxide wet etch process steps. This layer was patterned in hot phosphoric acid using a hard mask made of 0.25 µm thick LPCVD TEOS patterned with BOE. After silicon nitride layer patterning, 265 nm of LPCVD TEOS was deposited as a cladding layer for the waveguides. As a last step, the oxide layers were removed from the waveguide facets with BOE etch, and wafer was diced into chips.

Micrographs of some fabricated filters are shown in Fig. 8, together with details of a 85:15 MMI and of the slightly bent junction in a 95:05 double MMI.

## 6. CHARACTERIZATION AND DISCUSSION

We have initially characterized the filters over a broadband range from 1530 nm to 1570 nm wavelength, using a SLED and an optical spectrum analyzer (OSA) (see Fig. 9.a), to then realize that the limited 70 pm resolution of the OSA was too coarse to fully capture the spectral features of the filters. This is why we have also made some additional measurements with a narrow-line tunable laser and 5 pm tuning steps (see Fig. 9.b). In both cases we have used a circulator to collect the light back-reflected from the filters.

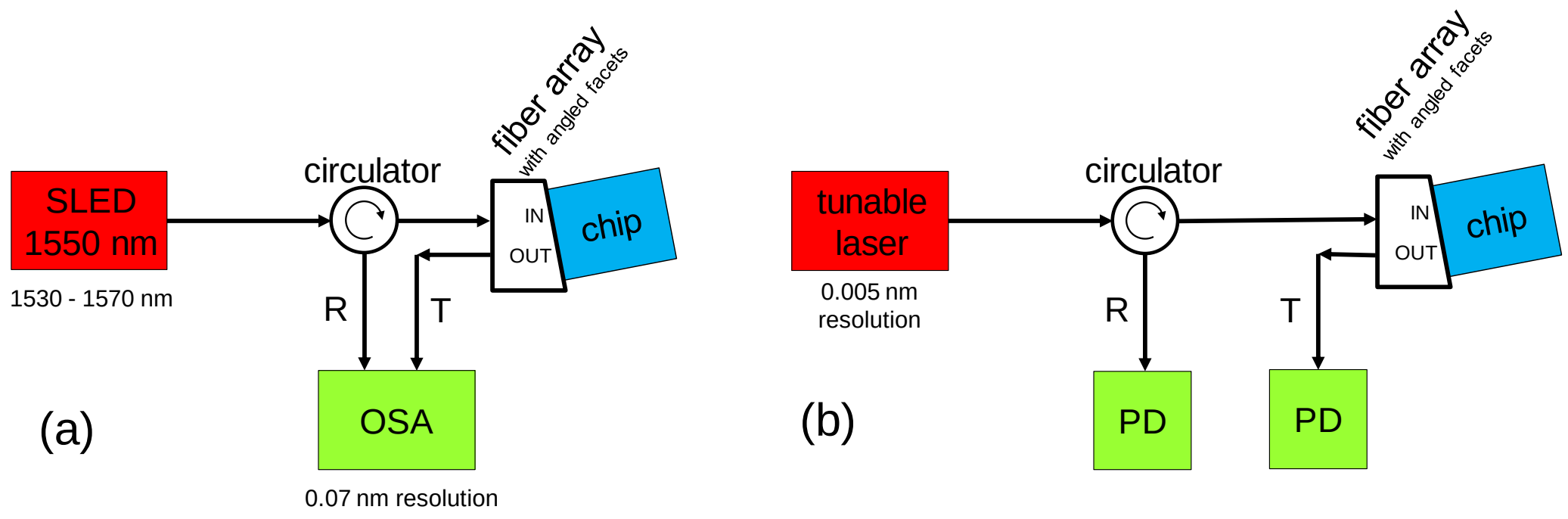

Fig. 9. Simulated spectral response of the chosen filters.

The low-resolution measurements are summarized in Fig. 10, not showing a proper second-order behavior in all cases. The four different filters have been measured from two different chips, and spectral features are not identical but consistent. High finesse filters show very high losses in the reflection spectrum, not to mention major spectral issues in the Type 3 implementation (with U-bends), most likely due to significant excitation of higher order modes.

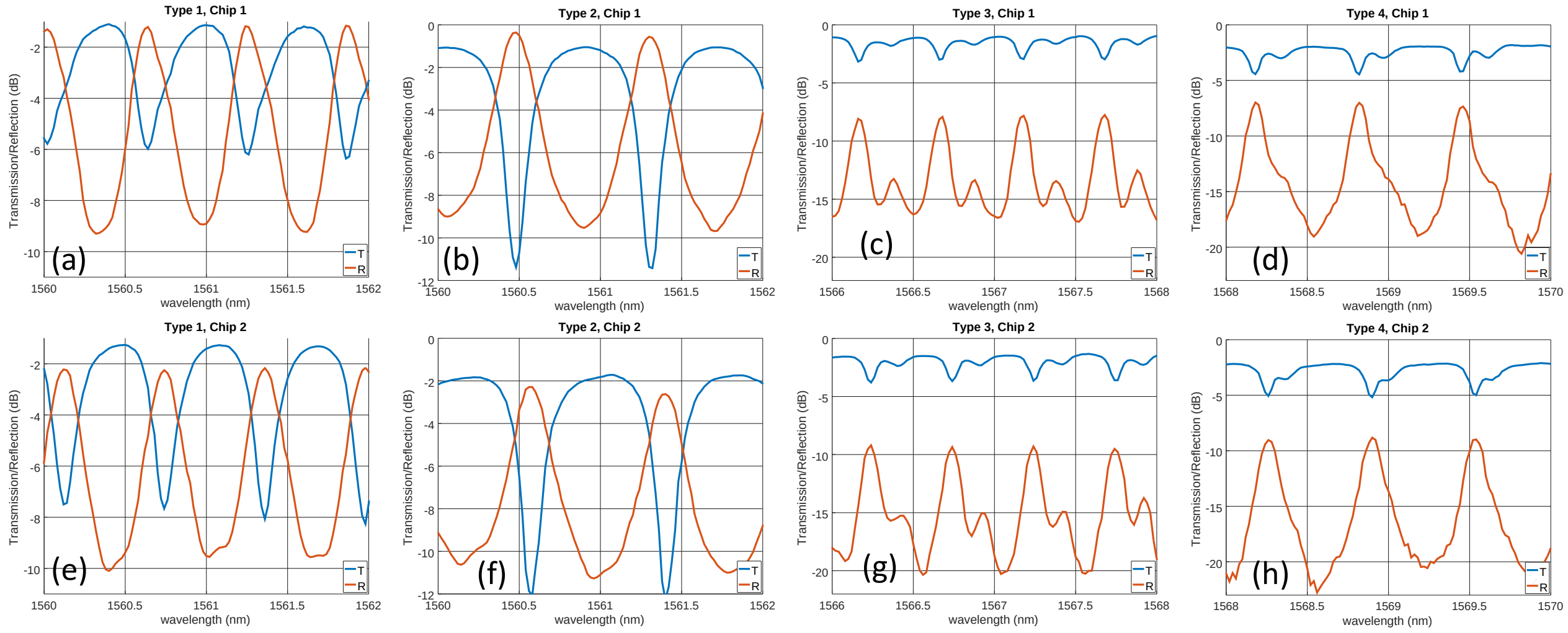

Fig. 10. Details of some resonant peaks measured with the SLED from two different fabricated chips: top row Chip 1, bottom row Chip 2.

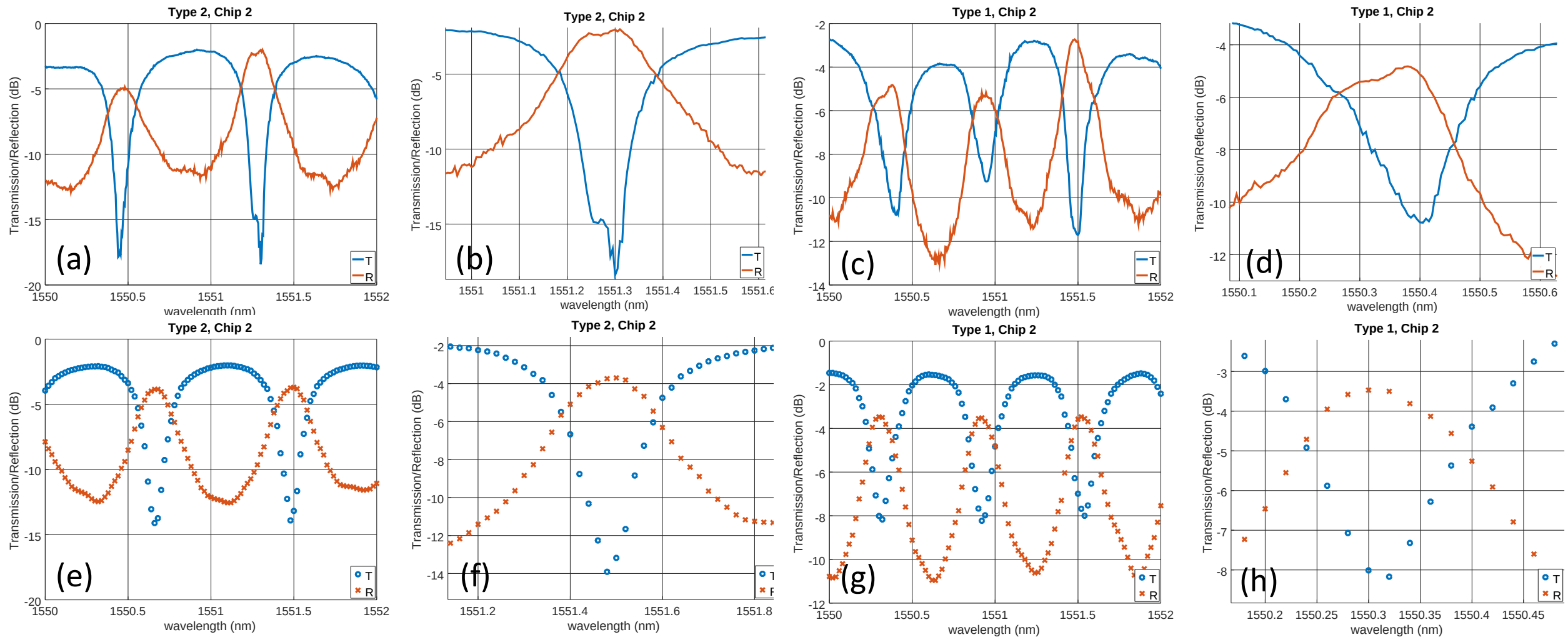

Fig. 11. Comparison of high resolution (top row) low-resolution (bottom row) spectra for Type 1 and Type 2 filters from Chip 2.

Type 1 and Type 2 filters from Chip 2 have been measured again using the tunable laser, and comparison shown in Fig. 11 clearly highlights the bad impact of convoluting the measured spectrum with the low resolution OSA. Instead the high-resolution measurements show second order responses, and in particular the right-hand side peak in Fig. 11.a, also zoomed in in Fig. 11.b, matches very nicely the target flat top design. On the other hand other peaks don't look as good as that one, probably due both to the impact of losses and beating of excited higher order modes.

By simulating the impact of losses on the spectral response, we have highlighted that, while the transmission response is independent of the chosen input port, as clearly seen by the symmetric structure of Eq. (1), this is not the case for the reflection, Eq. (2) lacking that symmetry. We show in Fig. 12 the simulated impact of 0.458 dB round-trip loss on the two designed filters, clearly highlighting this symmetry breaking.

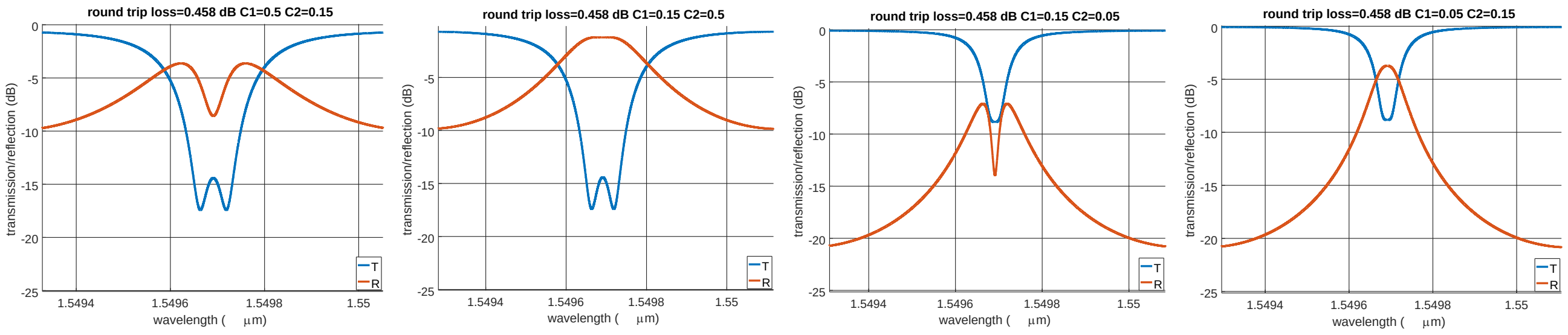

Fig. 12. Simulated spectral response of the chosen filters entering different ports assuming non-negligible loss.

## 7. CONCLUSIONS

We have introduced a new design of second-order filter based on a single ring resonator and implanted in our platform, by exploiting the unique properties of MMI splitters. The designed devices are based on different splitting ratios for the two splitters of the rings, targeting flat-top response instead of Fano resonances. The fabricated devices show second order behavior, but only in some cases the spectra match the design. Designs with high finesse suffer from high losses. We have also highlighted by simulation symmetry breaking of the reflected spectrum, depending on what port is used as an input.

## REFERENCES

[1] Madsen, C. K.., Zhao, J. H., Optical Filter Design and Analysis: A Signal Processing Approach, Wiley (1999).

[2] Bogaerts, W., De Heyn, P., Van Vaerenbergh, T., De Vos, K., Kumar Selvaraja, S., Claes, T., Dumon, P., Bienstman, P., Van Thourhout, D., et al., "Silicon microring resonators," Laser Photonics Rev. **6**(1), 47–73 (2012).

[3] Melloni, A.., Martinelli, M., "Synthesis of direct-coupled-resonators bandpass filters for WDM systems," J. Light. Technol. **20**(2), 296–303 (2002).

[4] Melloni, A., "Synthesis of a parallel-coupled ring-resonator filter," Opt. Lett. **26**(12), 917–919 (2001).

[5] Fan, S., Suh, W.., Joannopoulos, J. D., "Temporal coupled-mode theory for the Fano resonance in optical resonators," JOSA A **20**(3), 569–572 (2003).

[6] Xu, Q., Sandhu, S., Povinelli, M. L., Shakya, J., Fan, S.., Lipson, M., "Experimental Realization of an On-Chip All-Optical Analogue to Electromagnetically Induced Transparency," Phys. Rev. Lett. **96**(12), 123901 (2006).

[7] Hu, T.., et, al., "Tunable Fano resonances based on two-beam interference in microring resonator," Appl. Phys. Lett. **102**(1), 011112 (2013).

[8] Wang, K., Yu, C., Zhang, X., Xu, C., Zhang, Y.., Yuan, P., "Electromagnetically induced-transparency-like spectrum in an add/drop interferometer," Appl. Opt. **54**(6), 1285–1289 (2015).

[9] Soref, R. A., Schmidtchen, J.., Petermann, K., "Large single-mode rib waveguides in GeSi-Si and Si-on-SiO2," IEEE J. Quantum Electron. **27**(8), 1971–1974 (1991).

[10] Solehmainen, K., Aalto, T., Dekker, J., Kapulainen, M., Harjanne, M.., Heimala, P., "Development of multi-step processing in silicon-on-insulator for optical waveguide applications," J. Opt. Pure Appl. Opt. **8**(7), S455–S460 (2006).

[11] Aalto, T., Solehmainen, K., Harjanne, M., Kapulainen, M.., Heimala, P., "Low-loss converters between optical silicon waveguides of different sizes and types," IEEE Photonics Technol. Lett. **18**(5), 709–711 (2006).

[12] Aalto, T., Harjanne, M., Ylinen, S., Kapulainen, M., Vehmas, T.., Cherchi, M., "Total internal reflection mirrors with ultra-low losses in 3 µm thick SOI waveguides," Proc SPIE **9367**, 93670B–93670B–9 (2015).

[13] Cherchi, M., Ylinen, S., Harjanne, M., Kapulainen, M.., Aalto, T., "Dramatic size reduction of waveguide bends on a micron-scale silicon photonic platform," Opt. Express **21**(15), 17814–17823 (2013).

[14] Cherchi, M., Ylinen, S., Harjanne, M., Kapulainen, M., Vehmas, T.., Aalto, T., "The Euler bend: paving the way for high-density integration on micron-scale semiconductor platforms," Proc SPIE **8990**, 899004-899004–899007 (2014).

[15] Cherchi, M.., Aalto, T., "Bent optical waveguide," WO2014060648 A1 (2014).

[16] Cherchi, M., Ylinen, S., Harjanne, M., Kapulainen, M., Vehmas, T.., Aalto, T., "Unconstrained splitting ratios in compact double-MMI couplers," Opt. Express **22**(8), 9245–9253 (2014).

[17] Doménech, J. D., Fandiño, J. S., Gargallo, B.., Muñoz, P., "Arbitrary Coupling Ratio Multimode Interference Couplers in Silicon-on-Insulator," J. Light. Technol. **32**(14), 2536–2543 (2014).